\documentclass[aps,showpacs,preprintnumbers,amsmath, amssymb]{revtex4}

\usepackage{float}
\usepackage{graphics,epsfig}
\usepackage{graphicx}
\usepackage{dcolumn}
\usepackage{bm}

\begin{document}
\baselineskip=0.8 cm

\title{{\bf Holographic superconductors in quintessence AdS black hole spacetime}}
\author{Songbai Chen}
\email{csb3752@163.com} \affiliation{ Institute of Physics and
Department of Physics, Hunan Normal University,  Changsha, Hunan
410081, P. R. China \\ Key Laboratory of Low Dimensional Quantum
Structures \\ and Quantum Control of Ministry of Education, Hunan
Normal University, Changsha, Hunan 410081, P. R. China}
\author{Qiyuan Pan}
\email{panqiyuan@126.com}
\affiliation{ Institute of Physics and
Department of Physics, Hunan Normal University,  Changsha, Hunan
410081, P. R. China \\ Key Laboratory of Low Dimensional Quantum
Structures \\ and Quantum Control of Ministry of Education, Hunan
Normal University, Changsha, Hunan 410081, P. R. China}

\author{Jiliang Jing}
\email{jljing@hunnu.edu.cn}
 \affiliation{ Institute of Physics and
Department of Physics, Hunan Normal University,  Changsha, Hunan
410081, P. R. China \\ Key Laboratory of Low Dimensional Quantum
Structures \\ and Quantum Control of Ministry of Education, Hunan
Normal University, Changsha, Hunan 410081, P. R. China}

\vspace*{0.2cm}
\begin{abstract}
\baselineskip=0.6 cm
\begin{center}
{\bf Abstract}
\end{center}

We present a solution of Einstein equations describing a
$d$-dimensional planar quintessence AdS black hole which depends on
the state parameter $w_q$ of quintessence.  We investigate
holographic superconductors in this background and probe effects of
the state parameter $w_q$ on the critical temperature $T_c$, the
condensation formation and conductivity. The larger absolute value
of $w_q$ leads to the lower critical temperature $T_c$ and the
higher ratio between the gap frequency in conductivity to the
critical temperature for the condensates. Moreover, we also find
that for the scalar condensate there exists an additional constraint
condition originating from the quintessence
$(d-1)w_q+\lambda_{\pm}>0$ for the operators $\mathcal{O}_{\pm}$,
respectively. Our results show that the scalar condensation is
harder to form and the occurrence of holographic dual superconductor
needs the stronger coupling in the quintessence AdS black hole
spacetime.

\end{abstract}

 \maketitle
\newpage
\section{Introduction}

The anti-de Sitter/conformal field theory (AdS/CFT) correspondence
\cite{ads1} indicates that a string theory on asymptotically AdS
spacetimes can be related to a conformal field theory on the
boundary, which now is a powerful tool to study strongly coupled
phenomena in quantum field theory \cite{ads2,ads3,ads4}. In recent
years, this holographic correspondence has been employed to study
the strongly correlated condensed matter physics from the
gravitational dual \cite{Hs0,Hs01,Hs02,Hs03,Hsf0, Hsf01}. This dual
consists of a system with a black hole and a charged scalar field,
in which the black hole admits scalar hair at temperature lower than
a critical temperature, but does not possess scalar hair at higher
temperatures. According to the AdS/CFT correspondence, the emergence
of a hairy AdS black hole means the occurrence of a second order
phase transition from normal state to superconducting state which
brings the spontaneous U(1) symmetry breaking in the dual CFTs
\cite{Hsa1}. Due to the potential applications to the condensed
matter physics, the property of the holographic superconductor have
been studied extensively in the various gravity models
\cite{a0,a01,a1,a2,a3,a4,a401,a40,a42,a5,a6,a7,a8,a8d1,GT,a81,a9,a10,a100,
a11,a12,a13,a14,a15,a151,a17,GSJ,gx,bw1,Rgc}. The effects of
nonlinear electrodynamics on the holographic superconductor are also
studied in \cite{Jing1,Jing2,Jing3,wuj1}.

On the other hand, our Universe is undergoing an accelerated
expansion, which could be explained by the assumption that our
Universe is filled with dark energy. It is an exotic energy
component with negative pressure and constitutes about $72\%$ of
present total cosmic energy. The simplest interpretation of such a
dark energy is a cosmological constant with equation of state $w=-1$
\cite{1a}. Although the cosmological constant model is consistent
with observational data, at the fundamental level it fails to be
convincing. The vacuum energy density falls far below the value
predicted by any sensible quantum field theory, and it unavoidably
yields the coincidence problem, namely, why the dark energy and the
dark matter are comparable in size exactly right now. Thus, the
dynamical scalar fields, such as quintessence \cite{2a}, k-essence
\cite{3a} and phantom field \cite{4a}, have been put forth as an
alternative of dark energy. Models of dark energy differ with
respect to the size of the parameter $w$ namely, the relation
between the pressure and energy density of the dark energy.

Although dark energy has been firstly proposed to explain the
accelerated cosmological expansion, a lot of efforts have been
devoted to probing the effects of dark energy on the black hole
physics. Some attempts have been done to construct black hole solutions
with dark energy by using of dynamical scalar fields \cite{pbh}. However,
 it is not shown directly  the dependence of the structure of spacetime on
the state parameter of dark energy in these black holes. Actually, in the
 presence of a dynamical scalar field, the field equations of Einstein
 gravity are complicated and it is difficult to obtain the black hole solution with
the state parameter of dark energy. Kiselev \cite{Kiselev} adopted to a
phenomenological model of quintessence and obtained a static black hole
solution which depends on the state parameter $w_q$ of quintessence. In
this phenomenological model, quintessence is constructed by a unknown fluid
with the energy-momentum tensor satisfied the additive and linear conditions,
rather than by a dynamical scalar field as in the cosmology. For this special
phenomenological fluid, the nature is not clear, but the effective equation of
state $w_q$ is in a range $-1<w_q<0$.
Thus, the quintessence in this black hole solution is different from that in the
time-dependent cosmological evolution where it is encoded in the ``fifth-force" scalar field.
Subsequently,
we \cite{Sb1} extend Kiselev's work to the high dimensional case and find that
the first law is universal for the arbitrary state parameter $w_q$
of quintessence. The quasinormal modes and Hawking radiation of
these black holes surrounded by quintessence have been studied
entirely in \cite{Sb1,Sb2,gyx}.  These investigations could help us further
disclosing the relationship between dark energy and black hole.
The main purpose in this paper is to study the properties of
holographic superconductor in the quintessence AdS black hole
spacetime and to see what effect of the state parameter $w_q$ on the
critical temperature, the condensation formation and conductivity.

This paper is organized as follows. In Sec. II, we present the
metric describing a $d$-dimensional planar quintessence AdS black
hole and give the holographic dual of this black hole by introducing
a complex charged scalar field. In Sec. III, we apply the
Sturm-Liouville analytical \cite{GSJ} and numerical methods to
explore the relations between the critical temperature and the state
parameter $w_q$ of quintessence. In Sec. IV, we probe the effect of
the state parameter on the electrical conductivity of the charged
condensates.
Finally in the last section we will include our
conclusions.

\section{Holographic dual in the quintessence AdS black hole spacetime}

In this section, we present firstly the metric describing a
$d$-dimensional planar quintessence AdS black hole and construct the
holographic dual of the quintessence AdS black hole by introducing a
complex charged scalar field. The metric ansatz describing a
$d$-dimensional planar black hole can be taken as
\begin{eqnarray}
ds^2=-f(r)dt^2+\frac{1}{f(r)}dr^2+r^2dx_idx^i,\;\;\;\;\;i=1,2,...,d-2,\label{m1}
\end{eqnarray}
which satisfies Einstein's field equation
\begin{eqnarray}
R_{\mu\nu}-\frac{1}{2}g_{\mu\nu}R-\frac{(d-1)(d-2)}{2L^2}g_{\mu\nu}=8\pi
T_{\mu\nu}.\label{Eins}
\end{eqnarray}
$L$ is the radius of AdS and $T_{\mu\nu}$ is the energy-momentum
tensor for matter field. As in \cite{Kiselev}, one can construct the
nonzero components of the energy-momentum tensor for quintessence as
\begin{eqnarray}
&&T^{\;t}_{t}=T^{\;r}_{r}=-\rho_q,\nonumber\\
&&T^{\;x_1}_{x_1}=T^{\;x_2}_{x_2}=...=T^{\;x_{d-2}}_{x_{d-2}}=\frac{\rho_q}{d-2}\bigg[(d-1)w_q+1\bigg],
\label{enf}
\end{eqnarray}
where $w_q$ is the state parameter of quintessence.
Substituting
Eqs.(\ref{m1}) and (\ref{enf}) into Einstein's field equation
(\ref{Eins}), we can obtain
\begin{eqnarray}
&&\frac{(d-2)f'(r)}{2r}+\frac{(d-2)(d-3)f(r)}{2r^2}-\frac{(d-1)(d-2)}{2L^2}=-\rho_q,\nonumber\\
&&\frac{f''(r)}{2}+\frac{(d-3)f'(r)}{r}+\frac{(d-3)(d-4)}{2r^2}f(r)-\frac{(d-1)(d-2)}{2L^2}
=\frac{ \rho_q}{d-2}[(d-1)w_q+1],
\end{eqnarray}
which means that
\begin{eqnarray}
r^2f''(r)+[(d-1)w_q+2d-5]r
f'(r)+(d-3)[(d-1)w_q+d-3)]f(r)-\frac{(d-1)^2(w_q+1)r^2}{L^2}=0.
\end{eqnarray}
The general solution of the above equation has the form
\begin{eqnarray}
f(r)=\frac{r^2}{L^2}+\frac{c_1}{r^{d-3}}+\frac{c_2}{r^{(d-1)w_q+d-3}},\label{meq1}
\end{eqnarray}
where $c_1$ and $c_2$ are normalization factors. The energy density
$\rho_q$ for quintessence can be described by
\begin{eqnarray}
\rho_q=\frac{(d-1)(d-2)c_2w_q}{2r^{(d-1)(w_q+1)}}.\label{density}
\end{eqnarray}
In general, for the usual quintessence, the energy density $\rho_q$
is positive and the state parameter $w_q$ is negative, which means
the constant $c_2$ must be negative. When $w_q=-1$ and
$c_2=-\frac{1}{L^2}$, one can find that the energy density of
quintessence becomes $\rho_q=\frac{(d-1)(d-2)}{2L^2}$, which reduces
to the cosmological constant $\Lambda$. Thus, the cosmological
constant can be treated as a special kind of quintessence and the
study of quintessence could help us understand further the
cosmological constant. Moreover, we also find that one could
construct a type of anti-quintessence with the negative energy
density $\rho_q$ by setting the positive normalization factor $c_2$
and then a bare negative cosmological constant could be constructed
by setting that the positive $c_2=\frac{1}{L^2}$ and the state
parameter $w_q=-1$. The metric (\ref{m1}) with the function
(\ref{meq1}) describes an interesting set of planar black hole. As
$c_1=-M$ and $c_2=0$, it reduces to the usual Schwarzschild anti-de
Sitter black hole. When $c_1=-M$, $c_2=q^2$ and $w_q=(d-3)/(d-1)$,
it can reduce to the Reissner-Nordstr\"{o}m anti-de Sitter black
hole.

Here, we focus only on the case of the usual quintessence with the
positive energy density, i.e., $c_2<0$. Taking $c_1=0$ and $c_2=-M$,
we can obtain a metric function of the quintessence AdS black hole
spacetime
\begin{eqnarray}
f(r)=\frac{r^2}{L^2}-\frac{M}{r^{(d-1)w_q+d-3}}.\label{q1m}
\end{eqnarray}
The above choice of the normalization factors $c_1$ and $c_2$ is
very convenient for us to study the properties of holographic
superconductors. In order to study the properties of $s$-wave
holographic superconductor in the black hole spacetime (\ref{m1})
with the metric function (\ref{meq1}), we must rely on the numerical
method to solve the equation of motion of the gauge field and scalar
field. However, in the case with $c_1\neq 0$, we find that it is
hard to look for the numerical solution for the scalar field and the
gauge field with the standard algorithm. Fortunately, in the case
with $c_1=0$, we find that this difficulty can be avoided so that we
can study further the effect of the quintessence on the holographic
superconductor. Moreover, this choice also leads to that the metric
can be reduced to the usual planar Schwarzschild AdS black hole as
the state parametric of quintessence $w_q=0$. It is clear that the
metric depends on the state parametric  $w_q$ and the dimensional
number $d$. The radius of the black hole is
$r_H=(ML^2)^{1/(d-1)(w_q+1)}$. The Hawking temperature of the
quintessence AdS black hole spacetime is given by
\begin{eqnarray}
T_H=\frac{(d-1)(w_q+1)r_H}{4\pi L^2},
\end{eqnarray}
which is a function of the state parameter $w_q$ of quintessence. As
$w_q$ tends to $-1$, the metric (\ref{m1}) describes a pure AdS
spacetime rather than a black hole since the metric coefficient
$f(r)=(\frac{1}{L^2}-M)r^2$ in this limit. It could be understand by
a fact that when $w_q=-1$ quintessence becomes indistinguishable
from a cosmological constant and so the overall negative
cosmological constant proportional to $1/L^2$ has been renormalized
slightly by the subtraction of $M$. Moreover, the Smarr's formula
for the quintessence AdS black hole can be expressed as
\begin{eqnarray}
TS=\frac{1}{d-2}\Theta_LL+\frac{(d-1)w_q+d-3}{d-2}\Theta_{q}M,\label{rel1}
\end{eqnarray}
with
\begin{eqnarray}
\Theta_L=\frac{F(d)r^{d-1}_H}{L^3},\;\;\;\;\;\;\;\;\;\;
\Theta_q=\frac{F(d)}{2r^{(d-1)w_q}_H}.\label{relv}
\end{eqnarray}
Here we treat the AdS radius $L$ and the parameter $M$ as
variables. $F(d)$ is only a function of dimension $d$. The corresponding
differential form of the first law of thermodynamics is
\begin{eqnarray}
TdS=\Theta_LdL+\Theta_{q}dM.\label{rel2}
\end{eqnarray}
Comparing it with the formula of the first law for usual black holes, one
can find that the parameter $M$ can be interpreted as the mass of the black
hole only if $w_q=0$. When $w_q\neq0$, the energy from quintessence is the
product of the parameter $M$ and the corresponding generalized force $\Theta_{q}$,
i.e., $E_q=\frac{(d-1)w_q+d-3}{d-3}\Theta_qM$. Thus, in a sense, the parameter $M$
could be called as the ``quintessence charge". The entire understanding of
such ``quintessence charge" $M$ needs us to clarify the true nature of quintessence in future.

Let us now consider an electric field and a charged complex scalar
field coupled via a Lagrangian
\begin{eqnarray}
\mathcal{L}=-\frac{1}{4}F_{\mu\nu}F^{\mu\nu}-
|\nabla_{\mu}\psi-iA_{\mu}\psi|^2-m^2\psi^2,
\end{eqnarray}
where $\psi$ is a charged complex scalar field and $F^{\mu\nu}$ is
the strength of the Maxwell field $F=dA$. Adopting to the ansatz
\begin{eqnarray}
A_{\mu}=(\phi(r),\underbrace{0,0,...,0}_{d-1}),\;\;\;\;\psi=\psi(r),\label{psiform}
\end{eqnarray}
one can find that the equations of motion for the complex scalar
field $\psi$ and electrical scalar potential $\phi(r)$ can be
written as
\begin{eqnarray}
\psi^{''}+(\frac{f'}{f}+\frac{d-2}{r})\psi'
+\frac{\phi^2\psi}{f^2}-\frac{m^2\psi}{f}=0,\label{e1}
\end{eqnarray}
and
\begin{eqnarray}
\phi^{''}+\frac{d-2}{r}\phi' -\frac{2\psi^2}{f}\phi=0,\label{e2}
\end{eqnarray}
respectively, where we set $L^2=1$ and a prime denotes the
derivative with respect to $r$. Here, we must point that we ignore
the backreaction for simplification in the subsequent numerical
calculations. The equation of motion including backreaction becomes
more complicated (see in the Appendix). Since there is no any
directed coupling among the quintessence, the
 complex scalar field $\psi$ and electrical field $\phi(r)$,
the energy momentum tensor originating from quintessence does not
appear in the equations of motion (\ref{e1}) and (\ref{e2}).
However, quintessence affects indirectly the behaviors of the scalar
and gauge fields $\psi(r)$ and $\phi(r)$ by the function $f(r)$. The
regularity condition at the horizon gives the boundary conditions
\begin{eqnarray}
&&\phi(r_H)=0,\nonumber\\
&&\psi(r_H)=-\frac{(d-1)(w_q+1)r_H}{m^2}\psi'(r_H).
\end{eqnarray}
At the spatial infinite $r\rightarrow\infty$, the scalar filed
$\psi$ and the scalar potential $\phi$ behave like
\begin{eqnarray}
\psi=\frac{\psi_{-}}{r^{\lambda_-}}+\frac{\psi_{+}}{r^{\lambda_+}},\label{b1}
\end{eqnarray}
and
\begin{eqnarray}
\phi=\mu-\frac{\rho}{r^{d-3}},\label{b2}
\end{eqnarray}
with
\begin{eqnarray}
\lambda_{\pm}=\frac{1}{2}\bigg[(d-1)\pm\sqrt{(d-1)^2+4m^2}\bigg].\label{lam1}
\end{eqnarray}
The constants $\mu$ and $\rho$ can be interpreted as the chemical
potential and the charge density in the dual field theory, respectively.
The coefficients $\psi_{-}$ and $\psi_{+}$ correspond to the vacuum
expectation values of the condensate operator $\mathcal{O}$ dual to
the scalar field according to the AdS/CFT correspondence. As in
\cite{ads}, we can impose the boundary condition that either
$\psi_{-}$ or $\psi_{+}$ vanish, so that the theory is stable in the
asymptotic AdS region.

\section{Dependence of critical temperature on the state parameter of quintessence}

In this section we will use both the Sturm-Liouville analytical
\cite{GSJ} and numerical methods to probe the dependence of critical
temperature on the state parameter of quintessence.

Introducing a new coordinate $z=r_H/r$, the equations of motion
(\ref{e1}) and (\ref{e2}) can be rewritten as
\begin{eqnarray}
&&\psi^{''}+\bigg(\frac{f'}{f}-\frac{d-4}{z}\bigg)\psi'
+\frac{r^2_H}{z^4}\bigg(\frac{\phi^2\psi}{f^2}-\frac{m^2\psi}{f}\bigg)=0,\label{e3}\\
&&\phi^{''}-\frac{d-4}{z}\phi^{'}-\frac{2r^2_H\psi^2}{z^4f}\phi=0,\label{e4}
\end{eqnarray}
where a prime denotes the derivative with respect to $z$. As
$T\rightarrow T_c$, one can find the condensation tends to zero,
i.e., $\psi\rightarrow0$. This means that near the critical
temperature the electric field can be expressed as
\begin{eqnarray}
\phi=\xi r_H(1-z^{d-3}),\label{ps23}
\end{eqnarray}
where $\xi=\rho/r^2_H$. Near the boundary $z=0$, one can introduce a
trial function $F(z)$ which obeys
\begin{eqnarray}
\psi\sim\frac{\psi_i}{r^{\lambda_i}}\sim\langle\mathcal{O}_i
\rangle\frac{z^{\lambda_i}}{r^{\lambda_i}_H}F(z),
\end{eqnarray}
with subscript $i=(+,-)$. The trial function $F(z)$ should satisfy
the the boundary condition $F(0)=1$ and $F'(0)=0$. The equation of
motion for $F(z)$ reads
\begin{eqnarray}
F''(z)&+&\bigg\{\frac{2\lambda_i}{z}-\frac{2+[(d-1)w_q+d-3]z^{(d-1)(w_q+1)}}{z(1-z^{(d-1)(w_q+1)})}
+\frac{d-4}{z}\bigg\}F'(z)
+\bigg\{\frac{\lambda_i(\lambda_i-1)}{z^2}\nonumber\\&-&
\frac{\lambda_i}{z}\bigg[\frac{2+[(d-1)w_q+d-3]z^{(d-1)(w_q+1)}}{z(1-z^{(d-1)(w_q+1)})}+\frac{d-4}{z}\bigg]
+
\frac{\xi^2(1-z^{d-3})^2}{(1-z^{(d-1)(w_q+1)})^2}\nonumber\\&-&
\frac{m^2}{z^2(1-z^{(d-1)(w_q+1)})}\bigg\}F(z)=0.\label{Fz1}
\end{eqnarray}
Multiplying the above equation with the following function
\begin{eqnarray}
T(z)=z^{2\lambda_i-d+2}(z^{(d-1)(w_q+1)}-1),
\end{eqnarray}
we can rewritten Eq. (\ref{Fz1}) as
\begin{eqnarray}
[T(z)F'(z)]'-Q(z)F(z)+\xi^2P(z)F(z)=0,\label{Fz2}
\end{eqnarray}
with
\begin{eqnarray}
Q(z)&=&-T(z)\bigg\{\frac{\lambda_i(\lambda_i-1)}{z^2}
-\frac{\lambda_i}{z}\bigg[\frac{2+[(d-1)w_q+d-3]z^{(d-1)(w_q+1)}}{z(1-z^{(d-1)(w_q+1)})}+\frac{d-4}{z}\bigg]
-\frac{m^2}{z^2(1-z^{(d-1)(w_q+1)})}\bigg\}, \nonumber\\
P(z)&=&T(z)\frac{(1-z^{d-3})^2}{(1-z^{(d-1)(w_q+1)})^2}.
\end{eqnarray}
Making use of the Sturm-Liouville method, one can obtain the minimum
eigenvalue of $\xi^2$ in Eq. (\ref{Fz2}), which can be calculated by
\begin{eqnarray}
\xi^2=\frac{\int^{1}_{0}[T(z)F'(z)^2+Q(z)F(z)^2]dz}{\int^{1}_{0}P(z)F(z)^2dz}.
\label{xi1}
\end{eqnarray}
The trial function $F(z)$ satisfied its boundary condition can be
taken as $F(z)=1-az^2$, so $\xi^2$ can be explicitly written as
\begin{eqnarray}
\xi^2=\frac{s(w_q,\lambda_i,d,a)}{t(w_q,\lambda_i,d,a)},
\end{eqnarray}
with
\begin{eqnarray}
s(w_q,\lambda_i,a)&=&-\bigg[\frac{\lambda^2_i+\lambda_i(d-1)w_q-4}{(d-1)w_q+2\lambda_i+4}
+\frac{4}{2\lambda_i-d+5}\bigg]a^2+\frac{2\lambda_i[\lambda_i+(d-1)w_q]}{(d-1)w_q+2\lambda_i+2}a
-\frac{\lambda_i(\lambda_i+(d-1)w_q)}{(d-1)w_q+2\lambda_i},\nonumber\\
t(w_q,\lambda_i,a)&=&\frac{1}{(d-1)(w_q+1)}\bigg\{\bigg[\text{PolyGamma}
\bigg(0,\frac{2\lambda_i+d+1}{(d-1)(w_q+1)}\bigg)
-2\text{PolyGamma}\bigg(0,\frac{2(\lambda_i+2)}{(d-1)(w_q+1)}\bigg)\nonumber\\
&+&\text{PolyGamma}\bigg(0,\frac{2\lambda_i-d+7}{(d-1)(w_q+1)}\bigg)
 \bigg]a^2-2\bigg[\text{PolyGamma}\bigg(0,\frac{2\lambda_i-d+5}{(d-1)(w_q+1)}\bigg)
 \nonumber\\&-&2\text{PolyGamma}\bigg(0,\frac{2(\lambda_i+1)}{(d-1)(w_q+1)}\bigg)
 +\text{PolyGamma}\bigg(0,\frac{2\lambda_i+d-1}{(d-1)(w_q+1)}\bigg)\bigg]a\nonumber\\&+&
 \bigg[\text{PolyGamma}\bigg(0,\frac{2\lambda_i-d+3}{(d-1)(w_q+1)}\bigg)
-2\text{PolyGamma}\bigg(0,\frac{2\lambda_i}{(d-1)(w_q+1)}\bigg)\nonumber\\
&+&\text{PolyGamma}\bigg(0,\frac{2\lambda_i+d-3}{(d-1)(w_q+1)}\bigg)\bigg]\bigg\}.
\end{eqnarray}
For different values of $w_q$, $d$ and $\lambda_i$, we can obtain
the minimum value of $\xi^2$ with appropriate value of $a$. Combing
$T=\frac{3(w_q+1)r_H}{4\pi}$ and $\xi=\frac{\rho}{r^2_H}$, we can
obtain the form of the critical temperature $T_c$
\begin{eqnarray}
T_c=\gamma\rho^{\frac{1}{d-2}},
\end{eqnarray}
with the coefficient $\gamma=\frac{w_q+1}{4\pi\xi^{1/d-2}_{min}}$.
Thus, we can probe the effects of the state parameter $w_q$ on the
critical temperature $T_c$ through estimating $\xi_{min}$ by using
the Sturm-Liouville method.

In Tables (I) and (II), we list the analytical and numerical values
of critical temperature for different $w_q$ and $\lambda_i$ in the
four and five dimensional quintessence AdS black hole spacetimes,
which shows that the analytical values of $T_c$ from the
Sturm-Liouville method agree entirely with the exact numerical
results.
\begin{table}[ht]
\begin{tabular}[b]{|c|c|c|c|c|c|c|c|c|c|c|}
\hline\hline \;\;\;\;\;\; &
  \multicolumn{4}{c|}{}&\multicolumn{6}{c|}{}\\
  \;\;\;\;\;\;\;\;&
  \multicolumn{4}{c|}{$T_c/\rho^{1/2}$ for $\mathcal{O}_-$}&
  \multicolumn{6}{c|}{$T_c/\rho^{1/2}$ for $\mathcal{O}_+$}\\
 \hline&\multicolumn{2}{c|}{}&\multicolumn{2}{c|}{}&\multicolumn{2}{c|}{}
 &\multicolumn{2}{c|}{}&\multicolumn{2}{c|}{}\\
 \;\;\;\; $w_q$\;\;\;\;
  &
  \multicolumn{2}{c|}{$\lambda_-=1$}&\multicolumn{2}{c|}{$\lambda_-=\frac{3}{2}$}&
  \multicolumn{2}{c|}{$\lambda_+=3$}&\multicolumn{2}{c|}{$\lambda_+=2$}
  &\multicolumn{2}{c|}{$\lambda_+=\frac{3}{2}$}\\&
  \multicolumn{2}{c|}{$(m^2=-2)$}&\multicolumn{2}{c|}{$(m^2=-\frac{9}{4})$}&
  \multicolumn{2}{c|}{$(m^2=0$}&\multicolumn{2}{c|}{$(m^2=-2)$}&
  \multicolumn{2}{c|}{$(m^2=-\frac{9}{4})$}\\
 \hline&&&&&&&&&&\\
 &\tiny{Analytical}&\tiny{Numerical}&\tiny{Analytical}
 &\tiny{Numerical}&\tiny{Analytical}&\tiny{Numerical}
 &\tiny{Analytical}&\tiny{Numerical}&\tiny{Analytical}
 &\tiny{Numerical}\\
 \hline&&&&&&&&&&\\
 0&\;$0.2250$&\;\;$0.2255$&\;\;$0.1507$&\;\;$0.1517$
 &\;\;$0.0844$&\;\;$0.0867$&$0.1170$&$0.1184$&0.1507&$0.1517$\\
 \hline&&&&&&&&&&\\
 -0.2&\;$0.2057$&\;\;$0.2059$&\;\;$0.1294$&\;\;$0.1297$
 &\;\;$0.0705$&\;\;$0.0717$&0.0987&0.0992&0.1294&$0.1297$\\
 \hline&&&&&&&&&&\\
 -0.4&\;$-$&\;\;$-$&\;\;$0.1200$&\;\;$0.1200$
 &\;\;$0.0567$&\;\;$0.0571$&0.0823&0.0824&0.1200&$0.1200$\\
 \hline&&&&&&&&&&\\
 -0.6&\;$-$&\;\;$-$&\;\;$-$&\;\;$-$
 &\;\;$0.0427$&\;\;$0.0427$&0.0763&0.0764&$-$&$-$\\
 \hline&&&&&&&&&&\\
 -0.8& \;\;$-$&\;\;$-$&\;\;$-$&\;\;$-$
 &\;\;$0.0276$&\;\;$0.0277$&$-$&$-$&$-$&$-$\\
 \hline\hline
\end{tabular}
\caption{Variety of the critical temperature $T_c$ for the operators
$\mathcal{O}_-$ and $\mathcal{O}_+$ with different values of
$\lambda$ and $w_q$ for $d=4$.}
\end{table}
For the same $\lambda$, we find that the critical temperatures $T_c$
for the scalar operators $\mathcal{O}$ decrease with the absolute
value $w_q$, which means that the quintessence with the smaller
$w_q$ makes it harder for the scalar hair to be condensated in this
background. This could be explained by a fact that the quintessence
with the smaller $w_q$ possesses the stronger negative pressure
which yields that the scalar condensate is harder to be formed.
\begin{table}[ht]
\begin{tabular}[b]{|c|c|c|c|c|c|c|c|c|c|c|}
\hline\hline \;\;\;\;\;\; &
  \multicolumn{4}{c|}{}&\multicolumn{6}{c|}{}\\
  \;\;\;\;\;\;\;\;&
  \multicolumn{4}{c|}{$T_c/\rho^{1/3}$ for $\mathcal{O}_-$}&
  \multicolumn{6}{c|}{$T_c/\rho^{1/3}$ for $\mathcal{O}_+$}\\
 \hline&\multicolumn{2}{c|}{}&\multicolumn{2}{c|}{}&\multicolumn{2}{c|}{}
 &\multicolumn{2}{c|}{}&\multicolumn{2}{c|}{}\\
 \;\;\;\; $w_q$\;\;\;\;&
  \multicolumn{2}{c|}{$\lambda_-=2-\sqrt{2}/2$}&\multicolumn{2}{c|}{$\lambda_-=2$}&
  \multicolumn{2}{c|}{$\lambda_+=4$}&\multicolumn{2}{c|}{$\lambda_+=2(1+\sqrt{2})$}&
  \multicolumn{2}{c|}{$\lambda_+=2$}\\
 &
  \multicolumn{2}{c|}{$(m^2=-\frac{7}{2})$}&\multicolumn{2}{c|}{$(m^2=-4)$}&
  \multicolumn{2}{c|}{$m^2=0$}&\multicolumn{2}{c|}{$(m^2=-2)$}&\multicolumn{2}{c|}{$(m^2=-4)$}\\
 \hline&&&&&&&&&&\\
 &\tiny{Analytical}&\tiny{Numerical}&\tiny{Analytical}
 &\tiny{Numerical}&\tiny{Analytical}&\tiny{Numerical}
 &\tiny{Analytical}&\tiny{Numerical}&\tiny{Analytical}
 &\tiny{Numerical}\\
 \hline&&&&&&&&&&\\
 0&0.3605&0.3608&0.2518&$0.2526$&\;$0.1676$&\;\;$0.1705$&$0.1825$&\;\;$0.1847$&0.2518&$0.2526$\\
 \hline&&&&&&&&&&\\
 -0.2&0.3153&0.3154&0.2103&$0.2106$&\;$0.1378$&\;\;$0.1392$&$0.1502$&\;\;$0.1512$&0.2103&$0.2106$\\
 \hline&&&&&&&&&&\\
 -0.4&$-$&$-$&0.1807&$0.1808$&\;$0.1080$&\;\;$0.1084$&$0.1182$&\;\;$0.1185$&0.1807&$0.1808$\\
 \hline&&&&&&&&&&\\
 -0.6&$-$&$-$&$-$&$-$&\;$0.0755$&\;\;$0.0779$
 &0.0979&\;\;0.0979&$-$&$-$\\
 \hline&&&&&&&&&&\\
 -0.8&$-$&$-$&$-$&$-$&\;$0.0462$&\;\;$0.0463$
 &$0.0591$&\;\;$0.0591$&$-$&$-$\\
 \hline\hline
\end{tabular}
\caption{Variety of the critical temperature $T_c$ for the operators
$\mathcal{O}_-$ and $\mathcal{O}_+$ with different values of
$\lambda$ and $w_q$ for $d=5$.}
\end{table}
\begin{figure}[ht]
\begin{center}
\includegraphics[width=7cm]{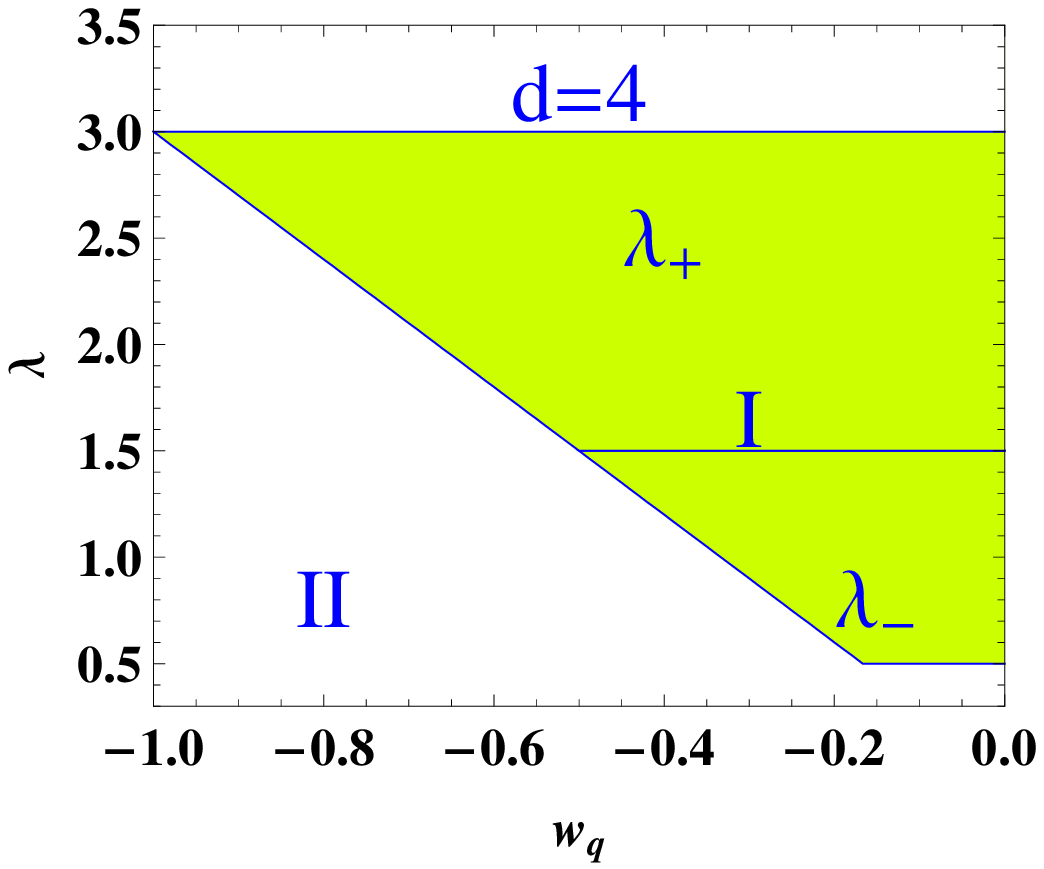}\;\;\;\;\;\includegraphics[width=7cm]{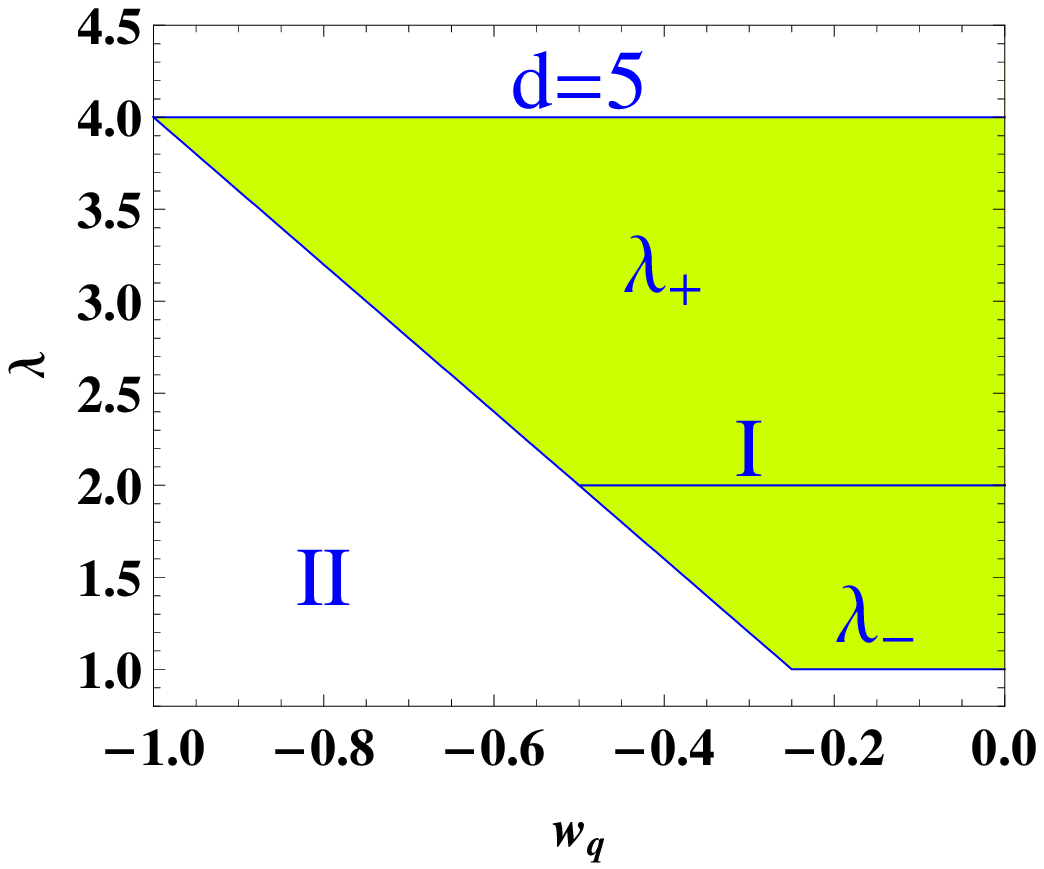}
\caption{The region I in the ($w_q$, $\lambda$) parameter space
denotes the allowable regions for the scalar condensates to be
formed. The left and right planes are for $d=4$ and $d=5$,
respectively. The middle horizontal line in the region I is the
boundary $\lambda=\frac{d-1}{2}$ between the allowable regions for
the operators $\mathcal{O}_-$ and $\mathcal{O}_+$. The top and
bottom horizontal lines correspond to the bounds $\lambda=d-1$ and
$\lambda=\frac{d-3}{2}$, respectively.}
\end{center}
\end{figure}
Moreover, we also find that the scalar condensate can be formed only
in the region
\begin{eqnarray}
w_q>-\frac{1}{2}\bigg[1+\sqrt{1+\frac{4m^2}{(d-1)^2}}\bigg],\;\;\;\text{and}\;\;\;
-\frac{(d-1)^2}{4}\leq m^2<0,
\end{eqnarray}
for the operator $\mathcal{O}_+$ and
\begin{eqnarray}
w_q>-\frac{1}{2}\bigg[1-\sqrt{1+\frac{4m^2}{(d-1)^2}}\bigg],\;\;\;\text{and}\;\;\;
-\frac{(d-1)^2}{4}\leq m^2<-\frac{(d-1)^2}{4}+1,
\end{eqnarray}
for the operator $\mathcal{O}_-$. This means that for the scalar
condensate there exists an additional constraint condition
originating from the quintessence
\begin{eqnarray}
(d-1)w_q+\lambda_{\pm}>0.\label{con36}
\end{eqnarray}
which is shown in tables (I), (II) and Fig. (1). The mathematical
reason is that the positive eigenvalues of Eq. (\ref{Fz2}) exists
only in the above regions. Beyond these regions, the eigenvalue
$\xi^2$ becomes negative, which results in that near the critical
temperature both of the electric field (\ref{ps23}) and the charge
density $\rho=\xi r^2_H$ are imaginary. In other words, there does
not exist any solution with physical meaning for the equations of
motion of the electric and scalar fields (\ref{e3}) and (\ref{e4}),
so the scalar condensate cannot be formed in this case. It could be
understand by a fact that with the smaller $w_q$ the strength of the
negative pressure of quintessence increases, but the critical
temperature $T_c$ decreases and the condensation becomes more
difficult to be formed. Beyond the region (\ref{con36}), the
negative pressure of quintessence becomes too strong to make the
scalar field condense. This  means that for a certain fixed
$\lambda_i$ beyond the region (\ref{con36}) the scalar condensate
could occur in the usual black hole spacetime, but now it does not
occur in the quintessence AdS black hole due to the negative
pressure of quintessence.  Moreover, the additional constraint
condition (\ref{con36}) tells us that for the quintessence with the
smaller $w_q$, the lower limit of $\lambda_{\pm}$  becomes larger
for the scalar field condensing. Moreover, we also find that the
region of existing scalar condensate for the operator
$\mathcal{O}_-$ is smaller than that for the operator
$\mathcal{O}_+$.
\begin{figure}[ht]
\begin{center}
\includegraphics[width=7cm]{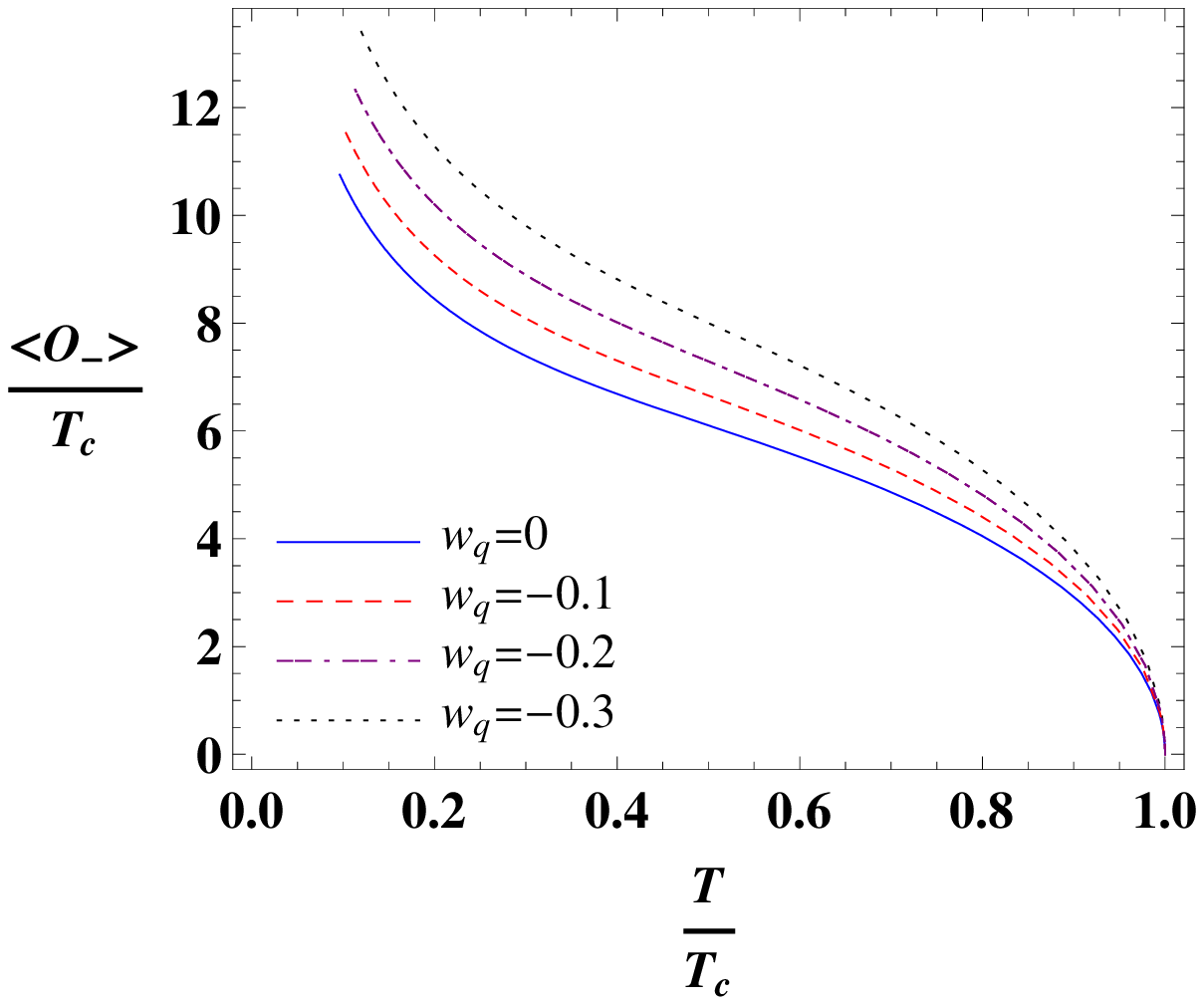}\;\;\;\;\;\includegraphics[width=7cm]{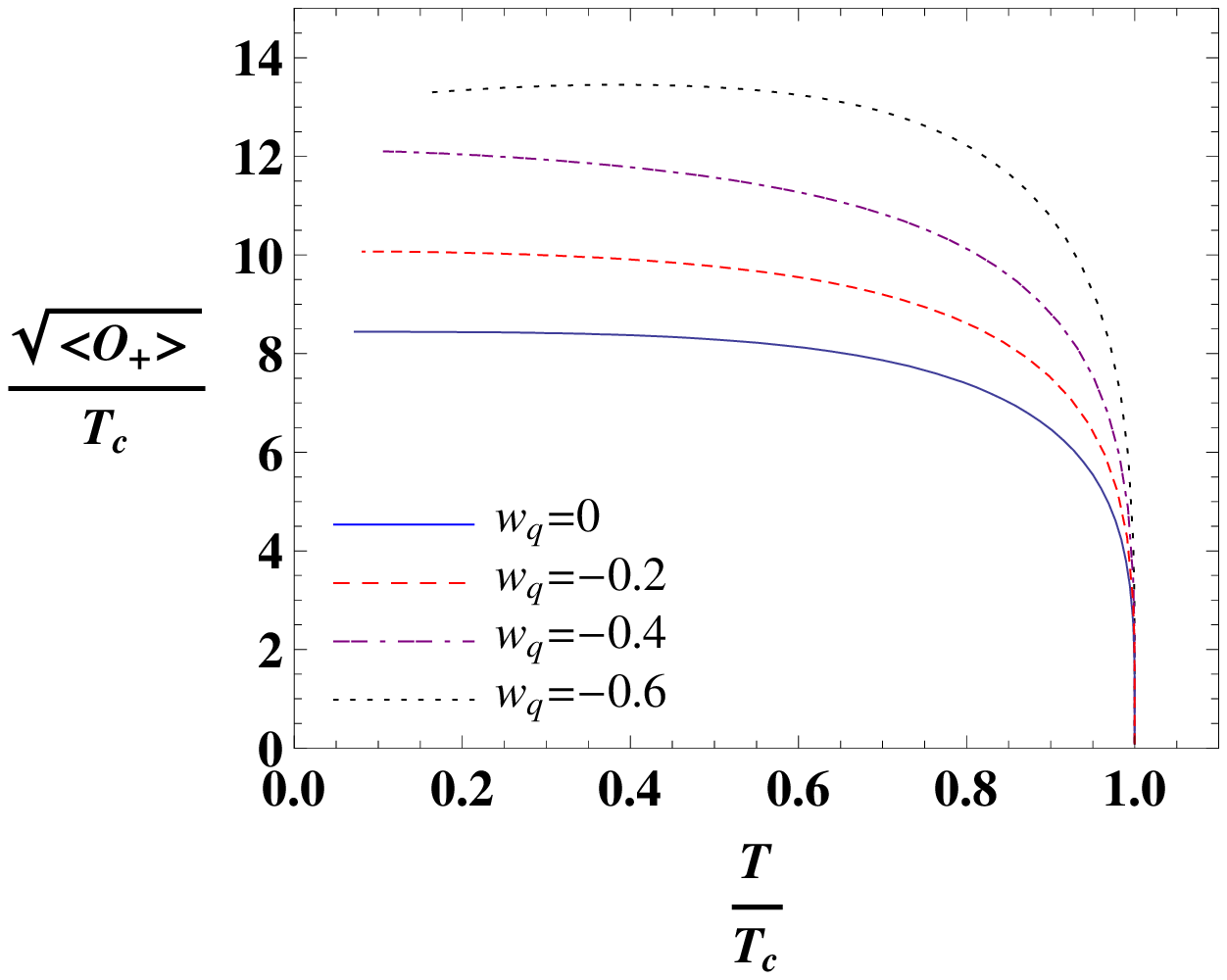}
\caption{The condensates of operators $\mathcal{O}_-$ (left) and
$\mathcal{O}_+$ (right) versus temperature for $m^2=-2$ in the case
$d=4$. The condensate is a function of temperature for various
values of state parameter $w_q$ in the allowable region (I).}
\end{center}
\end{figure}
\begin{figure}[ht]
\begin{center}
\includegraphics[width=7cm]{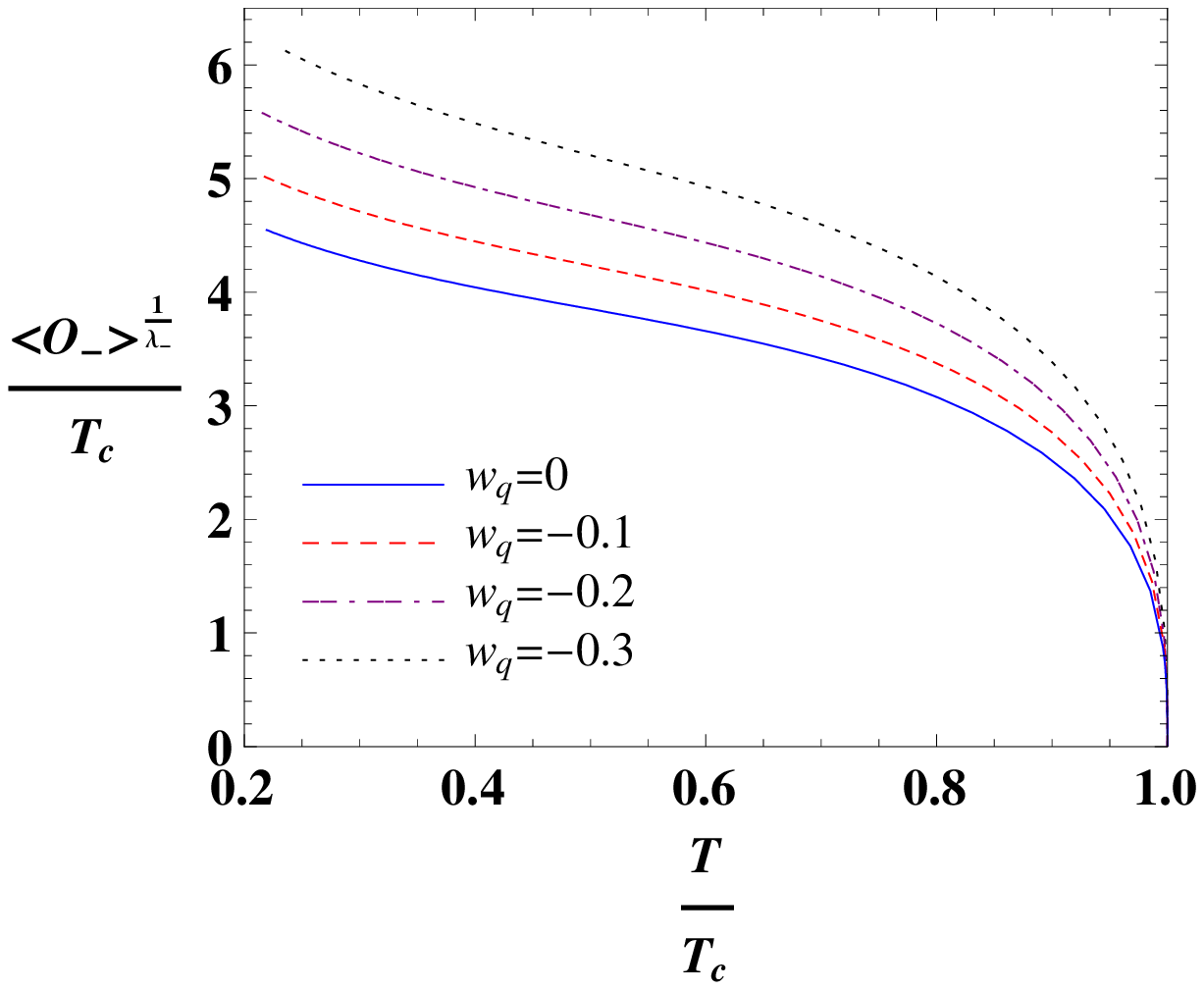}\;\;\;\;\;\includegraphics[width=7cm]{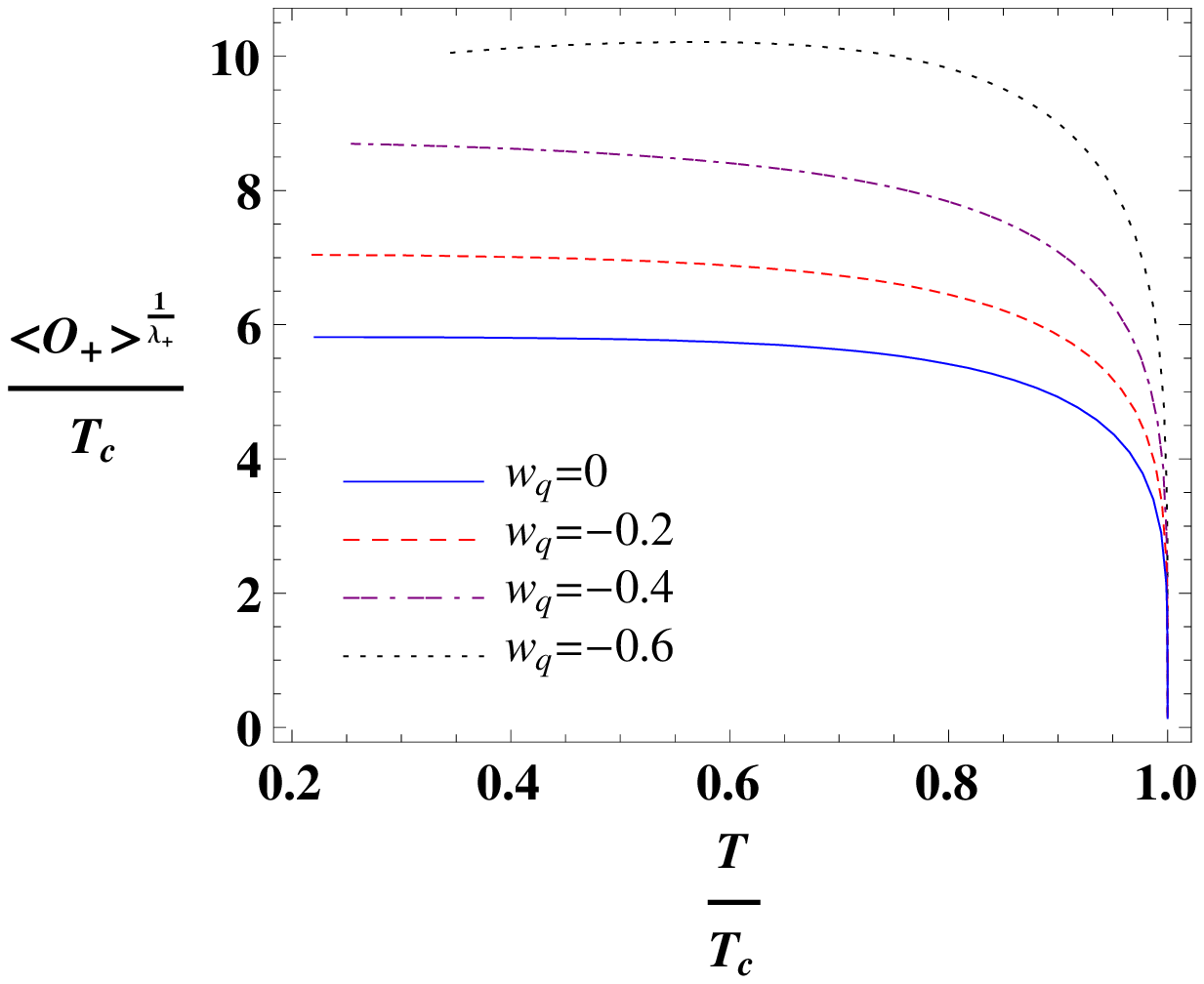}
\caption{The condensates of operators $\mathcal{O}_-$ (left) and
$\mathcal{O}_+$ (right) versus temperature for $m^2=-\frac{15}{4}$
in the case $d=5$. The condensate is a function of temperature for
various values of state parameter $w_q$ in the allowable region
(I).}
\end{center}
\end{figure}
In Fig. (2), we also plot the condensates of operators
$\mathcal{O}_-$ and $\mathcal{O}_+$ as a function of temperature
with the state parameter $w_q$ of quintessence in the allowable
region (I) for fixed $m^2=-2$ in the four dimensional quintessence
AdS black hole spacetime. The curves of $\mathcal{O}_-$ and
$\mathcal{O}_+$ for the non-zero $w_q$ have similar behavior to
those in the usual Schwarzschild AdS black hole spacetime with
$w_q=0$. Moreover, from Fig. (2), it is easy to find that the
condensation gap increases with the absolute value $w_q$. The
similar properties of the condensation gap with the absolute value
$w_q$ in the five dimensional black hole are shown in Fig. (3). These
imply that the condensation gap becomes larger and the scalar hair
is formed more difficultly in the quintessence AdS black hole
spacetime.

\section{The electrical conductivity}

In this section we will investigate the influence of the state
parameter $w_q$ of quintessence on the electrical conductivity.
Since the condensation gap and the critical temperature depend on
the state parameter $w_q$, it is natural for us to examine whether
the state parameter $w_q$ of quintessence will change the expected
universal relation in the gap frequency $\omega_g/T_c\approx8$
\cite{Hs01}. Following the standard procedure in \cite{Hs0,Hs01}, we
here adopt to a sinusoidal electromagnetic perturbation $\delta A_x
=A_x(r)e^{-i\omega t}$. Neglecting the backreaction of the
perturbational field on background metric, one can find the
electromagnetic perturbation obeys the first-order perturbational
equation
\begin{eqnarray}
A_x^{''}+\bigg(\frac{f'}{f}+\frac{d-4}{r}\bigg)A_x'
+\bigg[\frac{\omega^2}{f^2}-\frac{2\psi^2}{f}\bigg]A_x=0.\label{de}
\end{eqnarray}
\begin{figure}[ht]
\begin{center}
\includegraphics[width=6.8cm]{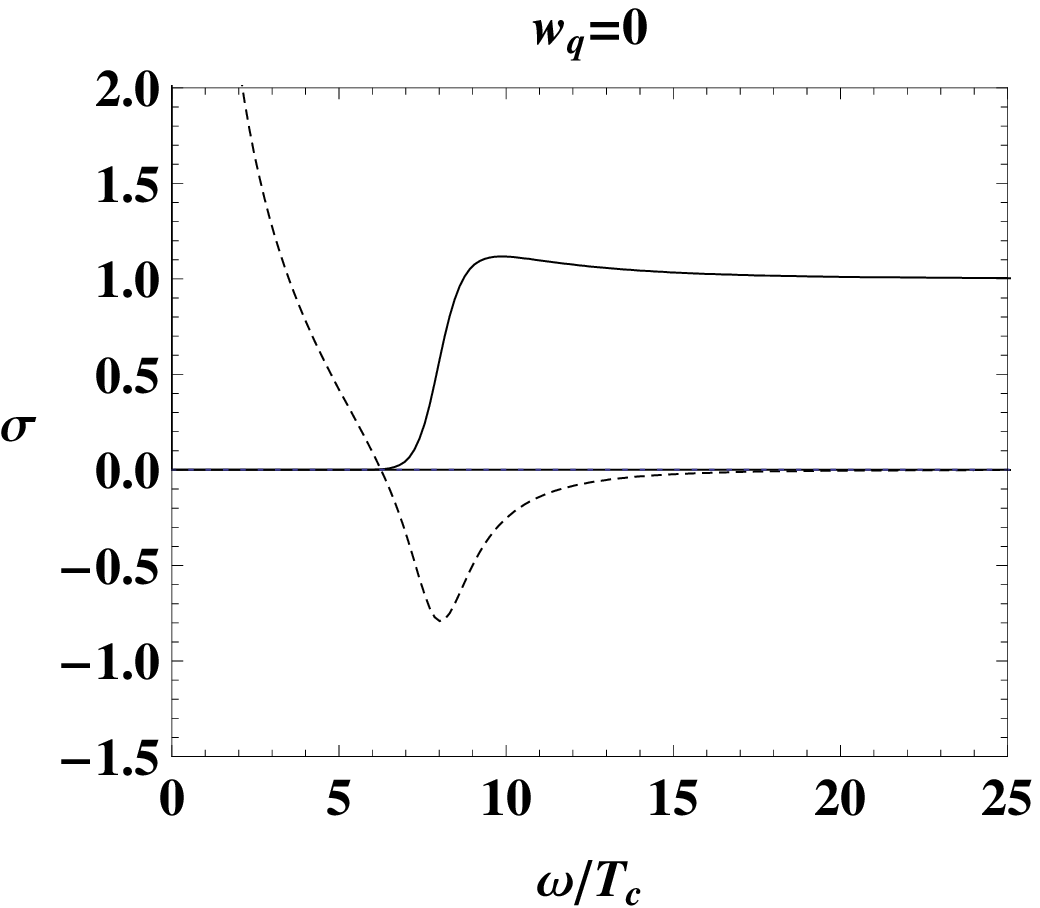}\;\;\;\includegraphics[width=6.8cm]{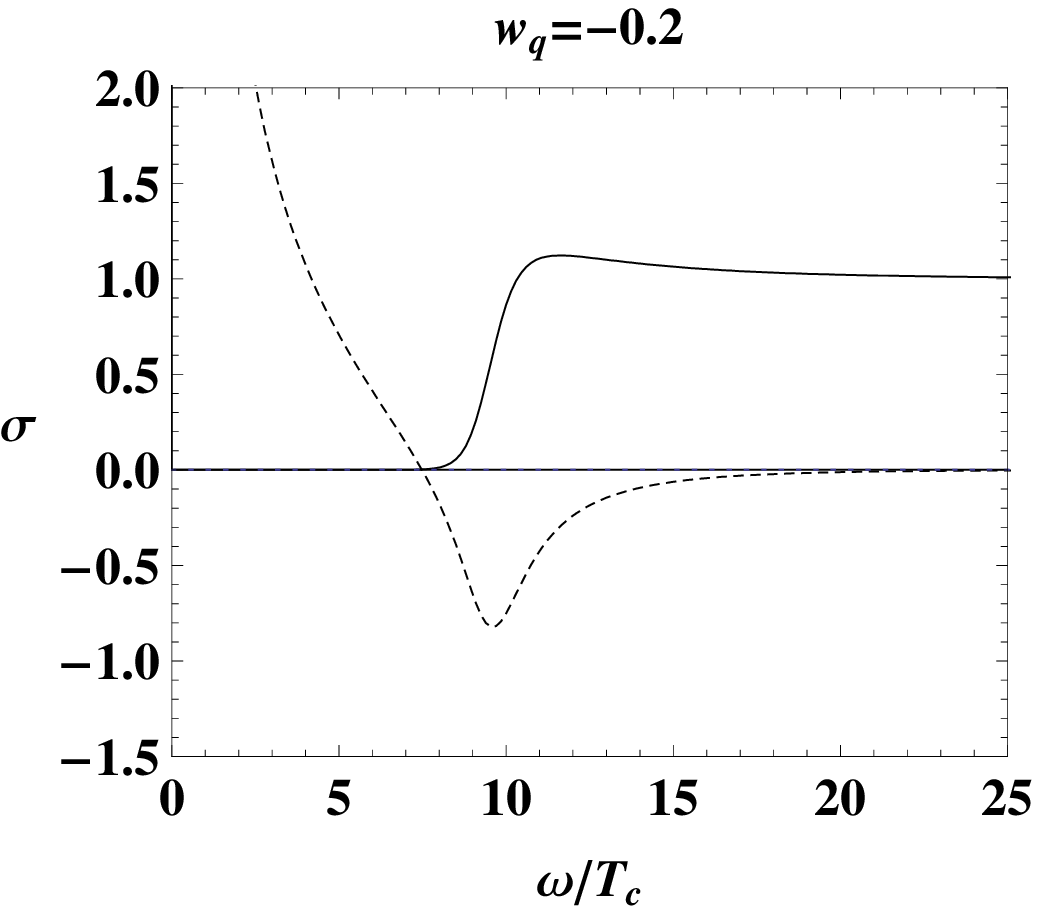}\\
\includegraphics[width=6.8cm]{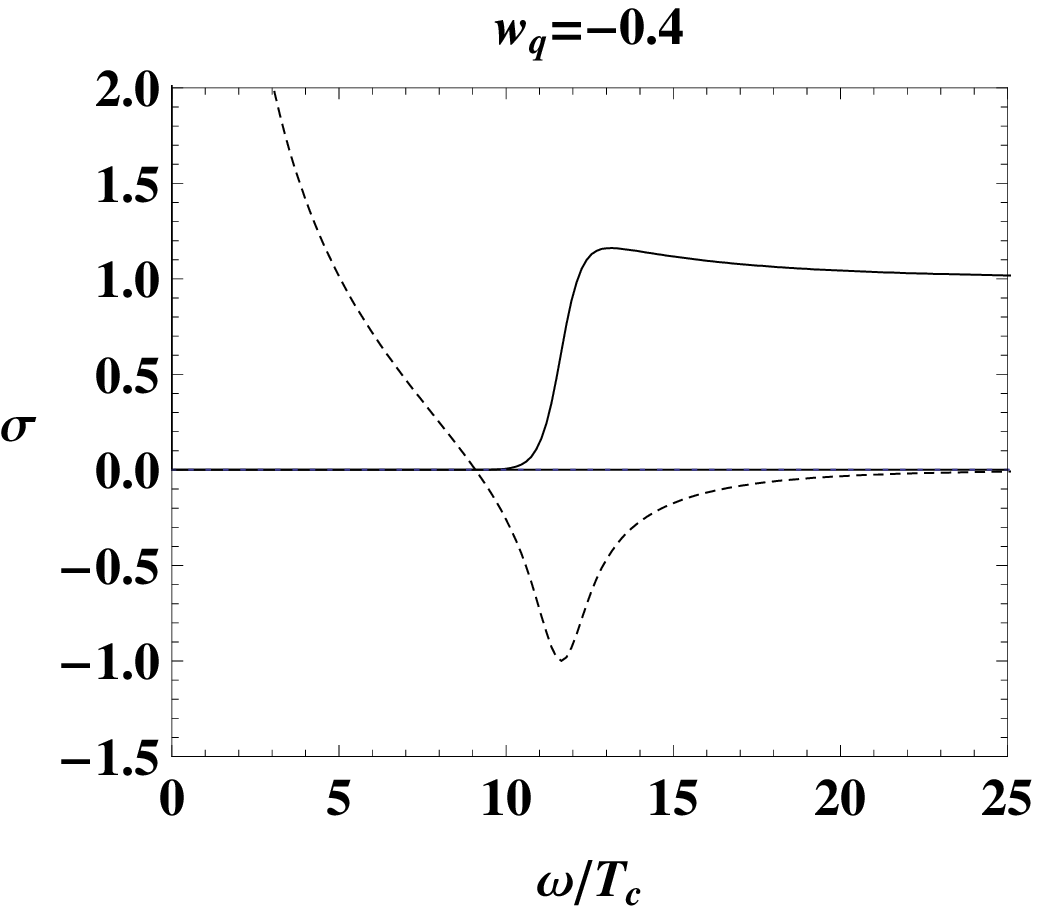}\;\;\;\includegraphics[width=6.8cm]{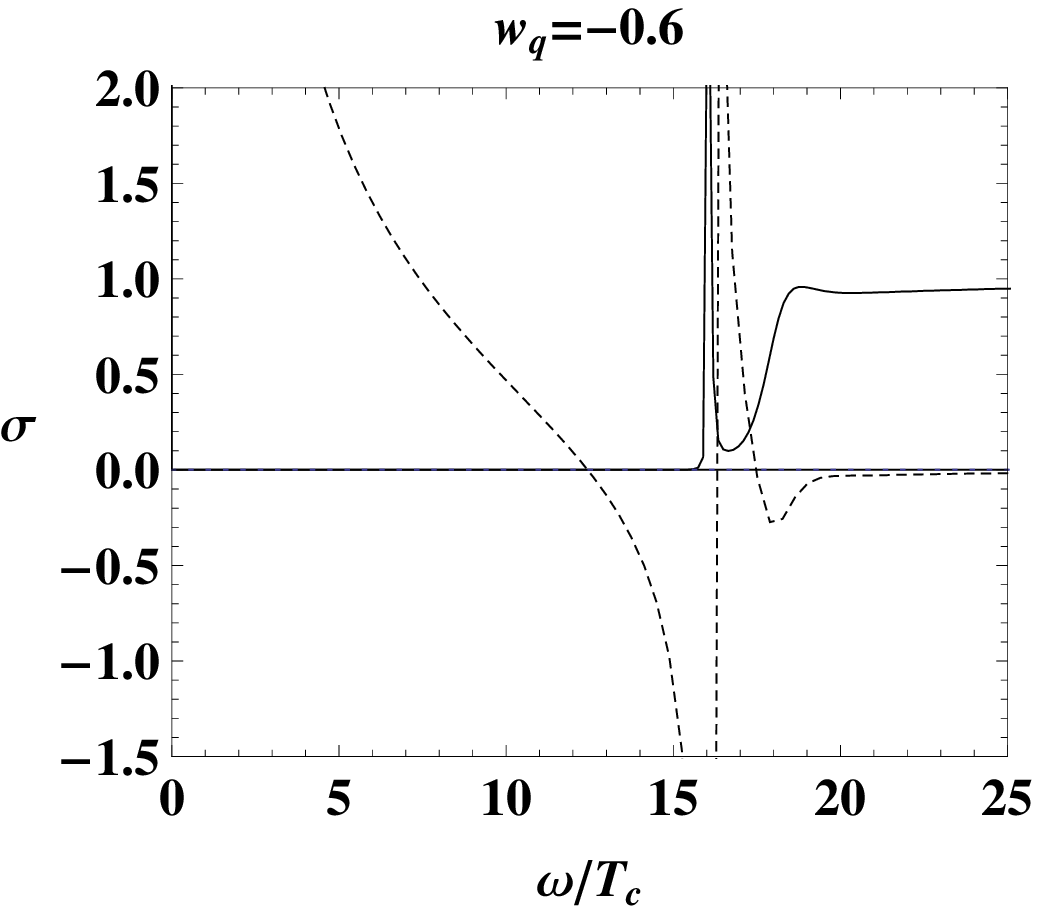}\\
\caption{The conductivity for operator $\mathcal{O}_+$ with
different values of $w_q$ in the four dimensional quintessence AdS
black hole spacetime. Each plot is at low temperatures, about
$T/T_c\approx 0.15$ and the fixed scalar mass $m^2=-2$. The solid
and dashed curves denote the real part $Re\sigma$ and the imaginary
part $Im\sigma$ of conductivity, respectively.}
\end{center}
\end{figure}
\begin{figure}[ht]
\begin{center}
\includegraphics[width=6.8cm]{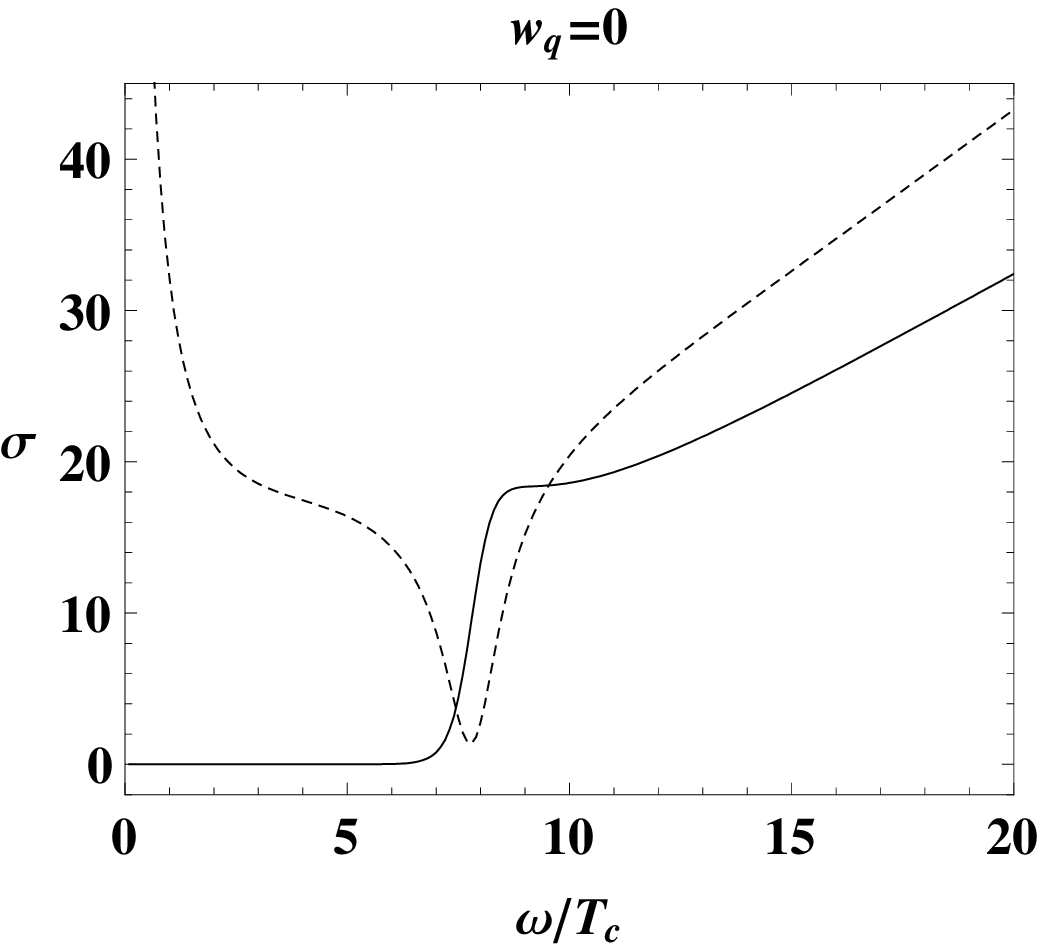}\;\;\;\includegraphics[width=6.8cm]{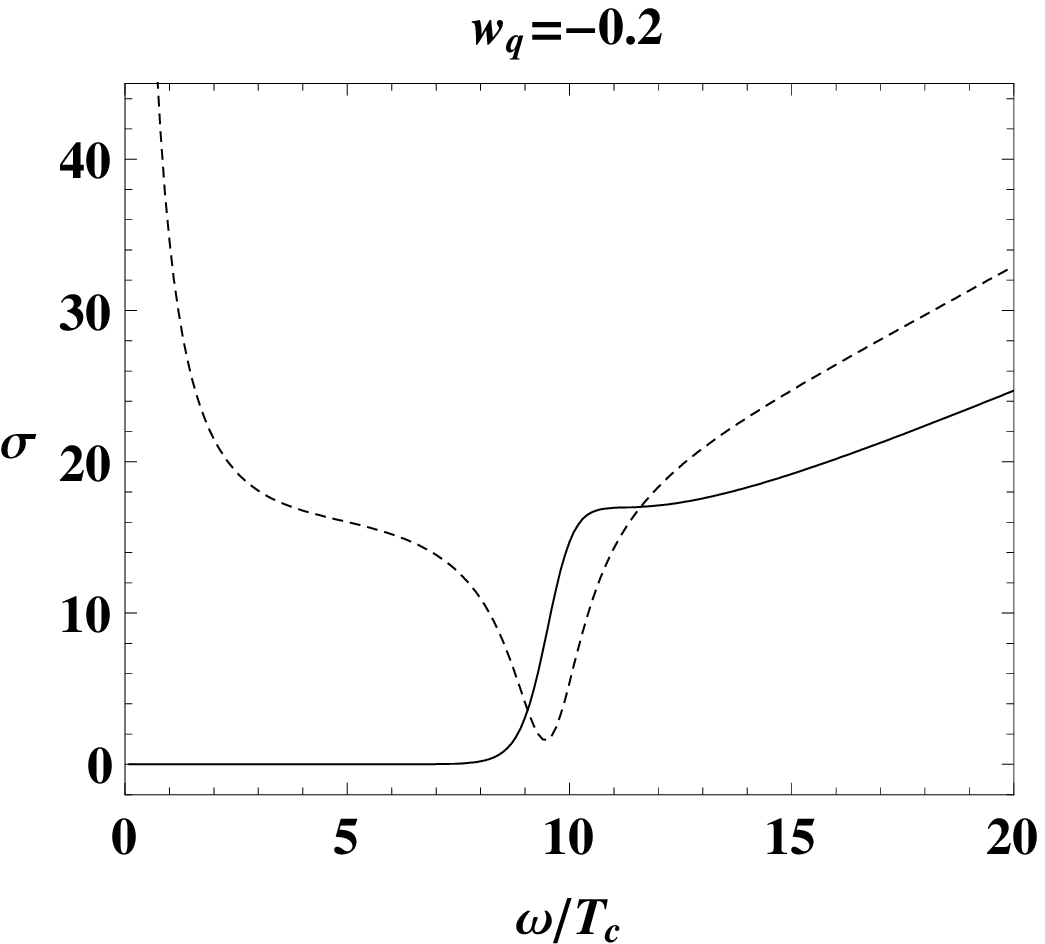}\\
\includegraphics[width=6.8cm]{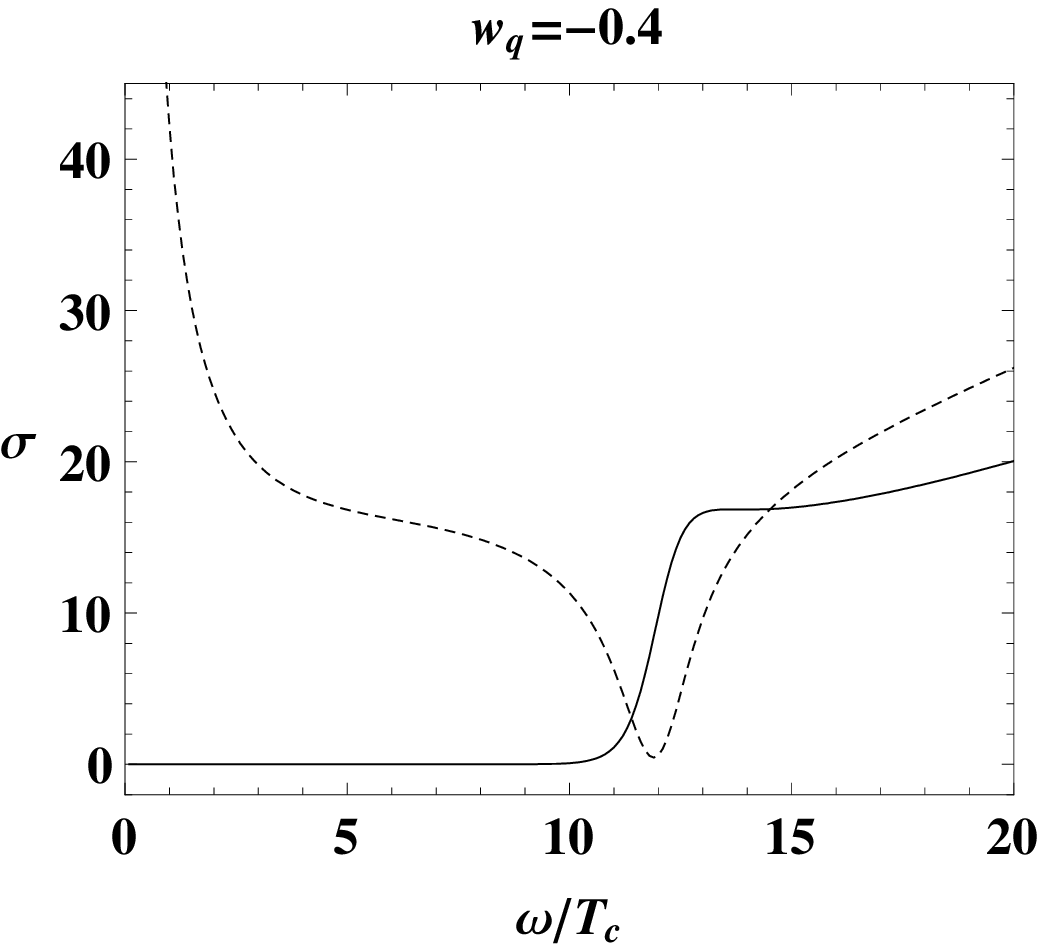}\;\;\;\includegraphics[width=6.8cm]{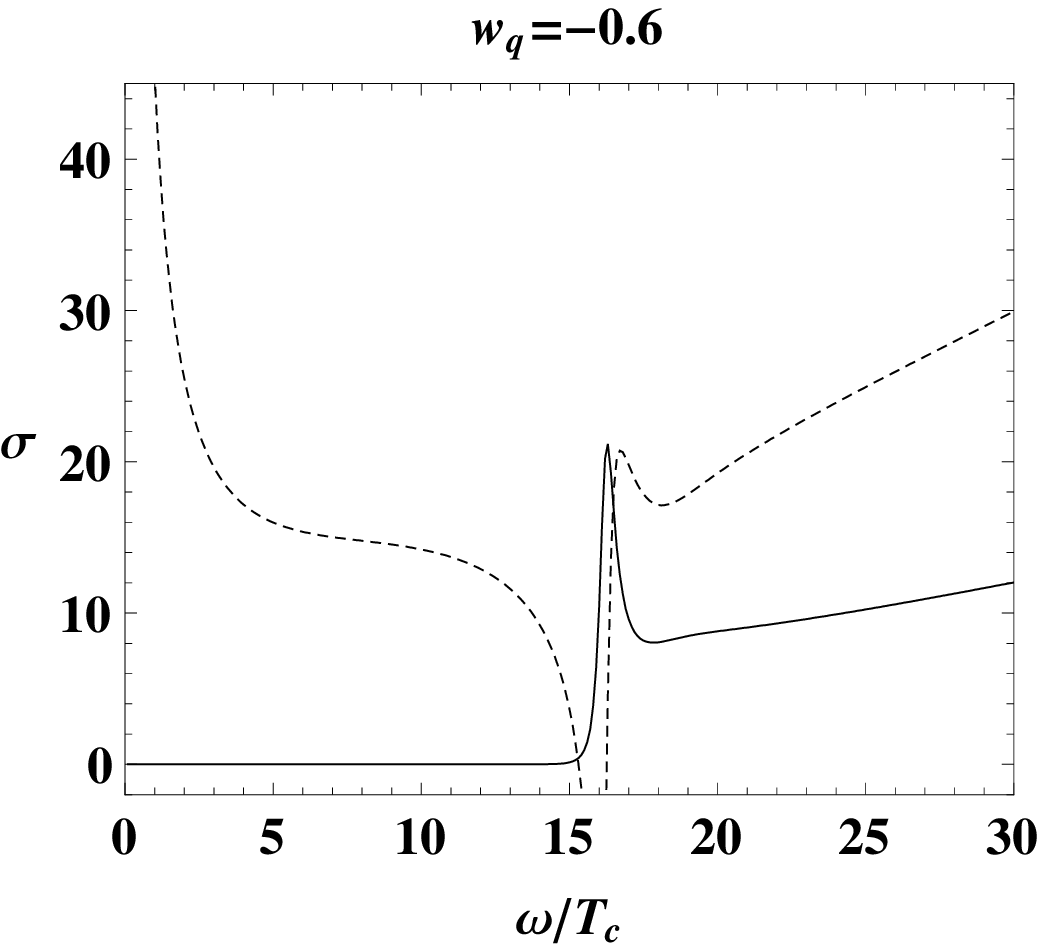}\\
\caption{The conductivity for operator $\mathcal{O}_+$ with
different values of $w_q$ in the five dimensional quintessence AdS
black hole spacetime. Each plot is at low temperatures, about
$T/T_c\approx 0.25$ and the fixed scalar mass $m^2=-3$. The solid
and dashed curves denote the real part $Re\sigma$ and the imaginary
part $Im\sigma$ of conductivity, respectively.}
\end{center}
\end{figure}
In order to avoid the complicated behavior in the gauge field
falloff in dimensions higher than five, we restrict our study to the
cases $d=4$ and $d=5$. The ingoing wave boundary condition near
horizon is given by
\begin{eqnarray}
A_x= f^{-\frac{i\omega }{(d-1)(w_q+1)r_H}},
\end{eqnarray}
and the general behavior of $A_x$ at the asymptotic AdS region can
be expressed as
\begin{eqnarray}
A_x=A^{(0)}_x+\frac{A^{(1)}_x}{r},
\end{eqnarray}
for the case $d=4$ and
\begin{eqnarray}
A_x=A^{(0)}_x+\frac{A^{(1)}_x}{r^2}+\frac{A^{(0)}_x\omega}{2}\frac{\log\Lambda
r}{r^2},
\end{eqnarray}
for the case $d=5$, respectively. From the AdS/CFT dictionary, one
can find \cite{Hs0} that $A^{(0)}_x$ and $A^{(1)}_x$ in the bulk
corresponds to the source and the expectation value for the current
on the CFT boundary, respectively. Thus, we can obtain the
conductivity of the dual superconductor by using the Ohm's
law\cite{Hs0}
\begin{eqnarray}
\sigma(\omega)=\frac{\langle J_x\rangle}{E_x}=-\frac{i\langle
J_x\rangle}{\omega A_x}=-i\frac{A^{(1)}_x}{\omega A^{(0)}_x},
\end{eqnarray}
for the case $d=4$ and
\begin{eqnarray}
\sigma(\omega)=\frac{\langle
J_x\rangle}{E_x}=-2i\frac{A^{(1)}_x}{\omega
A^{(0)}_x}+\frac{i\omega}{2},
\end{eqnarray}
for the case $d=5$, respectively. For different values of $w_q$, one
can obtain the conductivity by solving the Maxwell equation
numerically. We will focus on the case for the fixed scalar mass
$m^2=-2$ in our discussion. In Fig. (4) we plot the frequency
dependent conductivity for operator $\langle\mathcal{O}_+\rangle$
obtained by solving the Maxwell equation (\ref{de}) numerically for
$w_q=0$, $-0.2$, $-0.4$ and $-0.6$ at temperatures
$T/T_c\approx0.15$ in the four dimensional quintessence AdS black
hole spacetime. The solid and red dashed curves represent the real
part and imaginary part of the conductivity $\sigma(\omega)$
respectively. We find a gap in the conductivity with the gap
frequency $\omega_g$. With increase of the absolute value $w_q$, the
gap frequency $\omega_g$ becomes larger, which means that we have
larger deviations from the value $\omega_g/T_c\approx8$ for
increasing absolute value $w_q$ of quintessence. From Fig. (5), we
obtain the similar dependence of the conductivity on the state
parameter $w_q$ in the five dimensional case. Thus, the expected
universal relation in the gap frequency is really changed in the
quintessence black hole AdS spacetime, which is similar to the
effect of the Gauss-Bonnet coupling. Our result also show that the
occurrence of holographic superconductor needs the stronger coupling
in the quintessence AdS black hole spacetime.

\section{Summary}

In this paper we obtain an exact solution of Einstein equations
describing a $d$-dimensional planar quintessence AdS black hole
which depends on the state parameter $w_q$ of quintessence. It can
reduce to the usual Schwarzschild AdS black hole as the parameter
$w_q$ tends to zero. Applying the Sturm-Liouville analytical and
numerical methods, we investigate holographic superconductors in the
quintessence AdS black hole and probe effects of the state parameter
$w_q$ on the critical temperature $T_c$ and the condensation
formation and conductivity. The larger absolute value of $w_q$ leads
to the lower critical temperature $T_c$ and make the scalar operator
harder to form. Moreover, we found that the ratio between the gap
frequency $\omega_g$ in the conductivity and the critical
temperature $T_c$ becomes larger as the state parameter $w_q$
increases. This means that the occurrence of holographic
superconductor needs the stronger coupling in the quintessence AdS
black hole spacetime, which could be explained by a fact that
quintessence possesses negative pressure. Our result show that the
holographic superconductor share some similar features in the
quintessence and Gauss-Bonnet AdS black holes. Furthermore, we also
find that for the scalar condensate there exists an additional
constraint condition originating from the quintessence
$(d-1)w_q+\lambda_{\pm}>0$ for the operators $\mathcal{O}_{\pm}$.
Beyond the allowable regions, the negative pressure of quintessence
is so strong that the scalar condensate cannot be formed. For the
smaller $w_q$, the allowable region becomes more narrow. Moreover,
the region of existing scalar condensate for the operator
$\mathcal{O}_-$ is smaller than that for the operator
$\mathcal{O}_+$.

Finally, it is worth pointing that the dual field theoretical interpretation
of the quintessence in the metric (\ref{m1}) is a formidable puzzle. The main reason
is that the nature of quintessence in the metric (\ref{m1}) is still unclear at present
and its form of the Lagrangian $\mathcal{L}_q$ isn't available.
We leave this open issue in future. With the further investigations
of the exotic components, one could interpret entirely such a black hole with quintessence in the dual theory.

\appendix
\section{The equation of motion including backreaction}

In this appendix, we will present the equation of motion including backreaction. Considering
the backreaction from matter field, the metric ansatz for the $d$-dimensional planar black hole can be expressed as
\begin{eqnarray}
ds^2=-f(r)e^{-\chi(r)}dt^2+\frac{1}{f(r)}dr^2+r^2dx_idx^i,\;\;\;\;\;i=1,2,...,d-2.
\label{m1ap}
\end{eqnarray}
Substituting the metric (\ref{m1ap}) into the action
\begin{eqnarray}
S=\int d^dx\sqrt{-g}\bigg[\frac{1}{2\kappa^2}(R+2\Lambda+\mathcal{L}_q)-\frac{1}{4}F_{\mu\nu}F^{\mu\nu}-
|\nabla_{\mu}\psi-iA_{\mu}\psi|^2-m^2\psi^2\bigg],\label{act6}
\end{eqnarray}
and then making use of the variational principle, one can obtain the equation of motion
with backreaction
\begin{eqnarray}
\chi'+\frac{2r}{d-2}\frac{\partial \mathcal{L}_q }{\partial f}+\frac{4\kappa^2r}{d-2}
\bigg(\psi'^2+\frac{q^2e^{\chi}\phi^2\psi^2}{f^2}\bigg)=0,\label{ae1}\\
f'+\frac{d-3}{r}f-\frac{(d-1)r}{L^2}-\frac{2r}{d-2}\bigg(\mathcal{L}_q-2\frac{\partial
\mathcal{L}_q }{\partial \chi}\bigg)+\frac{2\kappa^2r}{d-2}\bigg[
m^2\psi^2+\frac{1}{2}e^{\chi}\phi'^2+f\bigg(\psi'^2+\frac{q^2e^{\chi}\phi^2\psi^2}{f^2}\bigg)
\bigg]=0,\label{ae21}\\
\phi''+\bigg(\frac{d-2}{r}+\frac{\chi'}{2}\bigg)\phi'-\frac{2q^2\psi^2}{f}\phi=0,\\
\psi''+\bigg(\frac{d-2}{r}-\frac{\chi'}{2}+\frac{f'}{f}\bigg)\psi'-\frac{m^2}{f}\psi
+\frac{q^2\phi^2}{f^2}\psi=0.\label{ae2}
\end{eqnarray}
Here the Lagrangian for the quintessence $\mathcal{L}_q$ could be a
function of $f$, $\chi$ and $w_q$. Since there is no any directed
coupling among the quintessence, the complex scalar field $\psi$ and
electrical field $\phi(r)$, the energy momentum tensor originating
from quintessence does not appear in the equations of motion for the
fields $\psi(r)$ and $\phi(r)$.

On the other hand, as in Ref.\cite{Kiselev}, the nonzero components
of the energy-momentum tensor of quintessence could be constructed
as a form (\ref{enf}) for the static spacetime (\ref{m1ap}) because
that quintessence meets the equation of state $p_q=w_q\rho_q$ in the
arbitrary spacetime. Here the pressure of quintessence $p_q$ is
defined by the average over the spatial part of the energy-momentum
tensor \cite{Kiselev}, i.e., $p_q=\langle T^{i}_{i}\rangle$,
($i=1,...,d-1$). According to the conservation of energy
$T^{\mu\nu}_{;\nu}=0$, one can find that in the spacetime
(\ref{m1ap}) the energy density $\rho_q$ of quintessence obeys
\begin{eqnarray}
\frac{d\rho_q}{dr}+(d-1)(w_q+1)\frac{\rho_q}{r}=0.
\end{eqnarray}
It can be solved analytically as
\begin{eqnarray}
\rho_q=\frac{s}{r^{(d-1)(w_q+1)}},
\end{eqnarray}
which is consistent with Eq.(\ref{density}) as we set the
integration constant $s=(d-1)(d-2)c_2w_q/2$. Therefore, one can find
that Einstein's field equation with backreaction can be written as
\begin{eqnarray}
f'+\frac{d-3}{r}f-\frac{(d-1)r}{L^2}-\frac{2\rho_qr}{d-2}
+\frac{2\kappa^2r}{d-2}\bigg[
m^2\psi^2+\frac{1}{2}e^{\chi}\phi'^2+f\bigg(\psi'^2+\frac{q^2e^{\chi}\phi^2\psi^2}{f^2}\bigg)
\bigg]=0,\label{newae1}\\
f'-f\chi'+\frac{d-3}{r}f-\frac{(d-1)r}{L^2}-\frac{2\rho_qr}{d-2}+\frac{2\kappa^2r}{d-2}\bigg[
m^2\psi^2+\frac{1}{2}e^{\chi}\phi'^2-f\bigg(\psi'^2+\frac{q^2e^{\chi}\phi^2\psi^2}{f^2}\bigg)
\bigg]=0.\label{newae21}
\end{eqnarray}
Thus, the equation of motion for $\chi$ can be expressed as
\begin{eqnarray}
\chi'+\frac{4\kappa^2r}{d-2}
\bigg(\psi'^2+\frac{q^2e^{\chi}\phi^2\psi^2}{f^2}\bigg)=0.
\end{eqnarray}
Comparing with Eq.(\ref{ae1}), one can find that $\frac{\partial
\mathcal{L}_q }{\partial f}=0$,  which means that the Lagrangian of
quintessence $\mathcal{L}_q$ is not an explicit function of $f$. It
is not surprising because that the Lagrangian of the electric field
also possesses the similar behavior. In this spacetime (\ref{m1ap}),
the Lagrangian of the electric field with the form (\ref{psiform})
can be written as
$\mathcal{L}_e=-\frac{1}{4}F_{\mu\nu}F^{\mu\nu}=\frac{1}{2}e^{\chi(r)}\phi'(r)^2$,
which is not an explicit function of $f$, but depends on $\chi(r)$.
Moreover, we find that the factor $\mathcal{L}_q-2\frac{\partial
\mathcal{L}_q}{\partial \chi} $ in Eq.(\ref{ae21}) can be replaced
by the energy density of quintessence $\rho_q$ and then
Eq.(\ref{ae21}) can be simplified further as Eq.(\ref{newae1}).
Although quintessence does not appear explicitly in the equations of
motion for the fields $\psi(r)$ and $\phi(r)$ Eq.(\ref{ae2}),
quintessence affects indirectly the behaviors of the scalar field
$\psi(r)$, the gauge field $\phi(r)$ and the quantity $\chi(r)$ by
the function $f(r)$.

\begin{acknowledgments}
This work was  partially supported by the National Natural Science Foundation of
China under Grant No.11275065, the NCET under Grant
No.10-0165, the PCSIRT under Grant No. IRT0964,  the Hunan Provincial Natural Science
Foundation of China (11JJ7001) and the construct
program of key disciplines in Hunan Province. J. Jing's work was
partially supported by the National Natural Science Foundation of
China under Grant Nos. 11175065, 10935013; 973 Program Grant No.
2010CB833004.
\end{acknowledgments}

\end{document}